# Non-classical dipoles in cold niobium clusters

Xiaoshan Xu, Shuangye Yin, Ramiro Moro, Anthony Liang,

John Bowlan, Walt A. de Heer

School of Physics, Georgia Institute of Technology, Atlanta, Ga 30332

**Abstract**: Electric deflections of niobium clusters in molecular beams show that they have permanent electric dipole moments at cryogenic temperatures but not higher temperatures, indicating that they are ferroelectric. Detailed analysis shows that the deflections cannot be explained in terms of a rotating classical dipole, as claimed by Anderson *et al*. The shapes of the deflected beam profiles and their field and temperature dependences indicates that the clusters can exist in two states, one with a dipole and the other without. Cluster with dipoles occupy lower energy states. Excitations from the lower states to the higher states can be induced by low fluence laser excitation. This causes the dipole to vanish.

## I. Introduction

After permanent electric dipoles were demonstrated in free niobium clusters in molecular beams,[1] their nature has been debated. Moro *et al*,[1] proposed that the dipoles were due to an unusual ferroelectric state since the dipoles vanished at moderate temperatures. The non-classical nature of the dipoles has been discussed in several studies.[2-9] However others claimed that the dipole moment is essentially classical, fixed to the cluster and actually do not vanish at room temperature. The reason they are not observed at higher temperature would be due to an averaging effect.[10-12] Here we show that the claimed averaging effect is incorrectly evaluated. More importantly, the experimental deflections cannot be explained in terms of a classical rotating dipole at any temperature. Niobium clusters represent non-classical dipoles in two senses. Firstly, for normal molecules and clusters that have a classical permanent dipole moments, the moments are relatively insensitive to temperature, in contrast to what we observe in niobium clusters. Secondly, the response of a niobium cluster is qualitatively different from that of a particle which has a dipole that is rigidly fixed to an axis of the particle. This non-classical behavior points towards a non-rigid coupling of the dipole moment.

Electric deflection experiments have been performed on cold niobium cluster beam produced in a cryogenic laser vaporization source. In contrast to the electric deflections of normal clusters,[13-15] the deflections of niobium clusters at low temperatures show several anomalies. In particular (i) the maximum deflection is linearly proportional to the applied electric field; (ii) the deflected beam profiles are asymmetric; (iii) at higher temperatures the spontaneous dipoles vanish and the deflections are due to the polarizability of the cluster. While (i) indicates that permanent electric dipoles are involved, the properties (ii) and (iii) need more careful examination.

Here we examine the deflection properties. In the simplest model one assumes that the dipoles are classical. Classical dipole are permanent and they are built into the cluster structure, typically because of an anisotropic charge distribution. Bertsch et al.[16]

theoretically examined the behavior of a thermally rotating cluster with a classical dipole moment in an inhomogeneous field. This classical rotating dipole picture was successfully applied by Dugourd *et al.*[17] to explain electric deflections of TiC$_{60}$. We take Bertsch's parameter free model as a starting point for the properties of rotating classical dipoles.

The force on a permanent dipole moment $\mu$ in an electric $E$ is $F=\mu\ dE/dz$. If the dipole moment rotates then the average force is reduced compared with the case where the dipole direction is aligned with the direction of the field gradient. Since free clusters rotate with an average rotational energy of 3/2 $k_BT$ it is clear that the average force in an electric field gradient be lower at finite temperature, compared with the force on a dipole that is aligned with the field. We show here that this thermal effect is orders of magnitude too small to explain the observed reduction at higher temperatures (in contradiction to the conclusions of Refs. 10 and 11). Moreover the predicted beam profile does not correspond to the experimentally observed profile. This shows that the observed deflections are non-classical.

## II. Deflection Experiment

The electric deflection experiment has been described in the Ref. 1 (see also Ref. 18). In brief, clusters are formed in a cryogenically laser vaporization cluster source that is cooled to temperature $T$: 15 K$\leq T \leq$300 K. The clusters thermally equilibrate in the source so that the resulting cluster beam is a frozen canonical ensemble that reflects the equilibrium population in the source as described in Ref. 19. The clusters enter the vacuum chamber that provides a collision-free environment for the clusters. The cluster beam is collimated and after traveling 1 m they deflect in an inhomogeneous electric field provided by specially shaped electrically charged plates in the deflection chamber. The geometry of the electric deflection plates is fixed, so that both the electric field $E$ and electric field gradient $dE/dz$ are proportional to voltage $V$ applied on the deflection plates.

The deflected clusters enter the position sensitive time of flight mass spectrometer where they are ionized by a pulse of UV light from an excimer laser. The mass spectrometer simultaneously measures the masses of the clusters as well as their positions in the detector chamber. In this way representative deflection profiles of all of the clusters in the beam are obtained.

The cluster deflection profiles obtained when the electric field is off represents the collimation function of the cluster beam. When the electric field is on, the clusters respond in various ways (Fig. 1). For example, for clusters like $Nb_{17}$, $Nb_{19}$, the profiles exhibit a rigid shift compared with the field off condition. The shift is found to be proportional to the square of the applied field strength. These clusters show normal polarizability behavior where a dipole moment proportional to the applied field is induced in the cluster. For other clusters (i.e. $Nb_{11}$, $Nb_{12}$, $Nb_{18}$ in Fig. 1.) the response is more complex. The shapes of the profiles are significantly altered; for these clusters the profiles become asymmetric and they are significantly broadened. Moreover, in contrast to normal clusters, the profile shapes are both field and temperature dependent. These features are discussed next.

**A. Field dependence**

The field dependence of the profile shape is illustrated in Fig. 2 that shows the beam profile of $Nb_{14}$ at several electric fields ($T$= 50 K). The profile consists of two components: a sharp peak that represents a rigid shift of the 0 field profile (dashed line) and an extended tail. The peak deflection is quadratic with the applied field and indicates a normal polarizable component. The extent of the tail is linear with the applied field and it represents a component with a permanent dipole moment.

**B. Temperature dependence**

Fig. 3 shows the temperature dependence of $Nb_{18}$ cluster as an example. At low temperatures, the deflected beam profiles is asymmetric, broadened and shifted. At $T$=300K, the profile is symmetric and rigidly shifted, indicating normal polarizable behavior. We show below that this shape change cannot be explained by the thermal averaging.

We next examine the deflections in detail and compare them with the classical dipole model.

## III. Comparison to the Classical Dipole Model

Bertsch *et al* [16] investigated the classical response of rotating cluster with a permanent dipole moment in a field. The dipole is classical in the sense that it is permanent (i.e. not field dependent) and it is rigidly fixed to an axis in the cluster. We use their classical dipole model (CDM) to calculate the deflections in an inhomogeneous field.

**A. Deflection profile simulation**

If the electric field is off, the beam profile observed is the collimation function $P_{off}(\delta)$ with a finite width, where $\delta$ is the deflection. If we define $I(\delta)$ as the beam profile observed when the collimation function is a delta function, with collimation $P_{off}(\delta)$ the observed profiles $P_{on}(\delta)$ is $P_{off}(\delta)$ convolved with $I(\delta)$. Note that if the electric field is turned off, $I(\delta)$ collapses to a delta function so that the deflection profile becomes the collimation function $P_{off}(\delta)$. Further note that when the deflections are much larger than the collimation width, $P_{on}(\delta)$ is approximately equal to $I(\delta)$. In principle one can obtain the experimental $I(\delta)$ by deconvolving the experimental $P_{off}(\delta)$ from the experimental $P_{on}(\delta)$. However, deconvolutions of this kind are difficult to perform on experimental data so that we proceed alternatively and use the classical dipole model to predict the experiment in order to obtain a more reliable comparison.

To simulate the deflection profiles from the theoretical response,[16] we follow the procedure described by Dugourd[17]. The steps are as follows.

(a) Calculate $<\cos\theta>$ for a cluster that starts from certain point of phase space, where $\cos\theta$

is the projection of dipole $\mu$ on field $E$.

(b) Repeat (a) to calculate the entire ensemble of clusters for the corresponding temperature[19] to get the distribution profile $I^{CDM}(<\cos\theta>)$.

(c) Use the formula:

$$= K(\mu dE/dz)/(mv^2) <\cos\theta> \quad (1)$$

to convert projection distribution profile $I^{CDM}(<\cos\theta>)$ to position distribution profile $I^{CDM}(\delta)$, $K$ is constant that depends on geometry of equipment, $m$ is the mass of cluster, and $v$ is the speed of the cluster.

(d) Convolve $I^{CDM}(\delta)$ with the collimation function $P_{off}(\delta)$ to get the calculated beam profile $P_{on}^{CDM}(\delta)$.

**B. TiC$_{60}$ the classical dipole case**

The deflections of the classical dipole molecule TiC$_{60}$ were measured by Dugourd et al,[17] who explained their deflections by applying the CDM above. Below we duplicate their calculations for that molecule.

The apparatus parameters relevant for Dugourd's experiment[17] as well for ours are shown in Table I. The parameters for the clusters are given in Table II. The required moments of inertia for TiC$_{60}$ are found from the rotational constants.[17]

In Fig. 4 we compare our calculations with those of Ref. 17 (which were obtained by digitizing Fig. 2 in Ref. 17). Our simulation is consistent with Dugourd's confirming that we correctly applied the model. As pointed out by Dugourd *et al*,[17] the calculated profiles reproduce the experimental results very well, which verifies that not only that TiC$_{60}$ is a classical dipole but also that the CDM model reliably predicts the experimental deflections of a classical dipole.

**C. Niobium clusters a non-classical dipole case**

The electric deflections of niobium clusters are clearly different than those for TiC$_{60}$ as the following analysis shows. Both the temperature dependence and the field dependence are non-classical.

The experimental beam profiles as well as the profiles calculated from the CDM at $T$=20K are shown in Fig. 5. The dipole moment of the cluster is determined from the extreme of the deflection profile $P_{on}$ in Fig. 5b. This shows that clusters deflect up to 2.2 mm at 5 kV. Note that the total beam intensity with the field off and with the field on are identical so that all of the deflected clusters are detected. However, using this value for the dipole moment gives a very poor fit in the CDM. The fit is even poorer for the 20 kV data, as shown in Fig. 5a. In fact, as can be seen, the model predicts rather symmetric deflections compared with the highly asymmetric deflections that are observed.

This situation does not improve at higher temperatures. Fig. 6a shows the calculated beam profile $P_{on}^{CDM}$ at $T=300$K and $V=20$kV. Again $P_{on}^{CDM}$ is broader than $P_{off}$ in contrast to the negligible broadening observed for $P_{on}$. This means that while the rotational averaging mechanism reduces the width of the deflected profiles, it does so to a far smaller degree than that we observe.

At higher temperatures, two factors affect cluster deflections. One is the rotation of the clusters, the other is their speed. The deflections are inversely proportional to the $v^2$(cf. Eqs. 1). Since $v^2$ is proportional to the beam temperature,[20] the deflections are inversely proportional to the temperature. This is a purely kinematic effect and larger deflections can be obtained by using a heavier carrier gas to reduce the speed. Figure 5a shows the 300 K deflections using He and Fig. 6b shows the deflections using Ar. Since the latter is 10 times heavier than the former, the deflections are increased by a factor of 10. Alternatively, replacing He with Ar has a similar effect at reducing the temperature by a factor of 10 compared with a He carried beam, at least as far as the kinematics are concerned. For both He and Ar carried beams, the CDM model fails to describe the deflections (see Fig. 6).

The failure of the CDM model for Nb clusters can be summarized as follows. (1)The CDM predicts symmetric profiles, which are not observed. (2) At high fields the CDM model predicts smaller peak intensities than observed. (3) The high temperature deflection profiles should be rather similar to those at low temperatures; the rotational averaging effect is rather similar in these two limits for the temperatures and fields used in the experiment. In contrast, in the experiment the differences between the low temperature and high temperature data are dramatic. The dipole is essentially absent at high temperatures.

The failure of the CDM model implies that the assumptions that go into this model do not apply to Nb clusters. Below we investigate reasons for the discrepancies. We find that there are two. Firstly, the profiles consist of two components, only one of which has a dipole moment that vanishes at high temperature. Secondly, the dipole moment is non-classical and appears not to be fixed to an axis in the cluster.

## IV. Two-component model

As shown in Fig. 2, the observed profiles suggest that they are composed of two components. The first component is normal and its deflections are characterized by an essentially rigid shift of the 0 field profile which is quadratic in the applied field. This is entirely consistent with the response due to the polarizability of the cluster. The second component produces the long tails, the extent of which varies linearly with applied field. Moreover, this component deflects in both the positive and in the negative directions giving appreciable broadening. The deflections of this component can be so large for sufficiently large fields, that the clusters are deflected out of the detector window. As indicated above, these properties are characteristic of a permanent dipole.

Most striking is that the intensity ratio of the two components is temperature dependent. The example above showed that at low temperatures the dipole component is large whereas at room temperature it is absent. The obvious explanation for this effect is that two states are involved, a lower energy state that has a permanent dipole moment and a higher energy state that is normal, so that with increasing temperature the low energy state is relatively less populated.

The implication of the two-component hypothesis is profound, and as pointed out in Ref. 1 it indicates that in these small clusters, two electronically very different states exist that are very close in energy. This is most certainly not expected since the energy scale for excited states is naturally set by the Fermi energy divided by the number of valence electrons in the cluster. By this counting, the energy scale for a 20 atom cluster for example should be of the order of 0.1 eV (i.e. 1000 K), which is at least two orders of magnitude greater than implied in the two-component model.

The two-component hypothesis is directly tested in the following laser heating experiment.

**A. Laser heating experiment**

The two-component model proposed above resulted from the failure of the CDM model to even qualitatively describe the observed deflections. We performed complementary laser-heating experiments to directly demonstrate that the temperature effect is not due to rotational averaging.

In these experiments (for details of this effect, see Refs. 21, 22 and 18), we illuminate the clusters in flight with a pulse of 500 nm laser light. This pulse is timed to illuminate clusters that are in transit from the source to the electric deflection plates. Hence the clusters are isolated from any thermal bath. Because of that, an absorbed photon can only affect the electronic and vibrational degrees of freedom but it specifically cannot alter the rotational state. Note that a single 500 nm photon will heat the cluster by about $9.7 \times 10^3$ /N K where N is the number of atoms in the cluster. Consequently, an absorbed photon heats a ground state $Nb_{30}$ cluster to about 300 K.

Results of the laser heating experiment are shown in Fig. 7. In this experiment, very cold $Nb_{28}$ clusters, produced in a *T*=20 K source were irradiated with the laser. Initially a broad peak is observed indicative of a large dipole component of this cluster. Low fluence laser heating causes the broad to collapse in a much narrower peak that is slightly shifted compared to the field free peak, which is typical for a normal (polarizable) cluster. The linearity of the effect with laser fluence was checked to insure that a single absorbed photon causes it. The effect was observed for all clusters in the beam that had dipoles. This experiment shows that rotational effects are not responsible for the vanishing of the dipole at higher temperature, but rather that they vanish due to increased internal energy as we originally claimed.[1]

**B. Two-component fit**

We next show that the deflection profiles can be well fit using two components: a normal polarizable component and a classical dipole component. As shown in Fig4, $P_{on}^{TC}$ are the profiles that result by assuming the two populations. Clearly, $P_{on}^{TC}$ matches $P_{on}$ much better than $P_{on}^{CDM}$, especially for the 20kV case, as shown in Fig. 5a. However, there still remains a significant discrepancy between $P_{on}^{TC}$ and $P_{on}$ which suggests that the CDM still not correctly describes the properties of the dipole part of the two components. Further investigations[1,19,4] show that this discrepancy is resolved if it is assumed that the dipole is not rigidly fixed the cluster's body. Note that the calculation of $P_{on}^{TC}$ finds for $Nb_{18}$ at this temperature about 50% population of polarizable component in both 5kV and 20kV conditions, indicating the population of the two components does not changes with electric field, which is consistent with the explanation of two states. The population changes with cluster size and beam temperature dramatically however. For 300K data shown in Fig. 6, obviously there is only polarizable component left because the $P_{on}$ basically is a rigid shift of $P_{off}$.

Anderson *et al*[10,11] performed density functional calculations on Nb clusters and concluded that these clusters had electric dipole moments which were remarkably close to those measured in Ref. 1. However their dipole moments are classical in both senses: they are rigidly fixed to the cluster and they are not temperature dependent (at least not on the temperature scales of the experiment).

Anderson *et al* concluded that the temperature effect reported in Ref. 1 was actually due to rotational averaging. In fact they used the CDM model above to prove their point. It turned out that their calculation in Ref. 10 was flawed, since even in 0 field their profiles $P^{CDM}(<cos\theta>)$ were asymmetric. In Ref. 11, the calculation was improved but still not correct because the profiles have rising tails for high fields. This anomalous result was probably obtained because Anderson *et al* determined the ensemble of the clusters by using their energies in the electric field. However we know the cluster ensemble is defined in the thermal equilibrium conditions in the source where they are formed and that changes in the electric field which is applied to the isolated clusters in the beam are adiabatic.[16,17] According to Ref. 16, the intensity always diminishes at the edge of the profile. In any case, the beam profile calculations in Refs. 10 and 11 are not realistic.

Their conclusion that the profile $P(<cos\theta>)$ for a classical dipole should be asymmetric when clusters rotate slowly is correct however only in relatively high fields or at very low temperatures. For our experimental conditions, the effect is negligible. For a cluster with electric dipole moment $\mu$=2 Debye at *T*=20K and *E*=80kV/cm, the ratio $\mu E/k_B T$=0.19, which still belongs to low field regime,[16] therefore the asymmetry of the profiles $P(<cos\theta>)$ is insignificant especially after it is convolved with $P_{off}$ in order to simulate the experimental profile as shown above. In any case, the asymmetry in the experimental profiles is beyond doubt caused by (at least) two components.

In summary we have investigated in detail whether the Nb cluster deflection data can be explained in terms of the single component classical dipole model. We conclude that the classical dipole moment consistently fails to reproduce observed beam profiles for all experimental conditions. Two components are required in order to correspond with experiment. The temperature dependence of the population ratio of the two components indicates that the dipole component represents a low energy state and the polarizable component represents a high energy state. This is directly verified by examining the effect of low fluence laser excitation that converts the dipole state to the polarizable state. The fact that clusters in the ground state have electric dipoles and slightly excited clusters do not is extremely significant, since there is no priori reason to expect low temperature ferroelectricity in such small clusters.

TABLE I: Parameters of apparatus. The calibration of the inhomogenous electric field is given by three parameters: $V_0$, $E_0$ and $dE_0/dz$, where the $E_0$ and $dE_0/dz$ are the electric field and field gradient respectively when the voltage applied on the deflection plates is $V_0$. One can find out the electric field and its gradient using the linear proportionality between $V$, $E$ and $dE/dz$.

|  | Dugourd's Setup | Our Setup |
|---|---|---|
| $K$ (m$^2$) | 0.165 | 0.277 |
| $E_0$ (V/m) | 1.63x10$^7$ | 8.00x10$^6$ |
| $dE_0/dz$ (V/m$^2$) | 2.82x10$^9$ | 3.25x10$^9$ |
| $V_0$ (V) | 2.7x10$^4$ | 2.0x10$^4$ |

TABLE II: Parameters of clusters. $I_1$ and $I_2$ are the principle moment of inertia of the clusters. For Nb$_{18}$, spherical shape and bulk density are assumed.

|  | TiC$_{60}$ | Nb$_{18}$ |
|---|---|---|
| $I_1$ (kgm$^2$) | 1.12x10$^{-39}$ | 2.04x10$^{-43}$ |
| $I_2$ (kgm$^2$) | 9.65x10$^{-40}$ | 2.04x10$^{-43}$ |
| $\mu$ (Debye) | 8.1 | 1.2 |
| $m$ (amu) | 768.47 | 1672.4 |

**Captions**

FIG. 1(color online). Representative beam profiles of Nb clusters at $T$=20K and $V$=20kV, where $V$ is the voltage applied on deflection plates that generate the inhomogeneous electric filed $E$. The zero field profile represents the collimation function, which dependences only on the geometry of the apparatus.

FIG. 2(color online). Beam profiles of Nb$_{14}$ at $T$=50K and several electric fields. For each profile we can distinguish a main peak and a tail. (b) The deflections of the main peaks and the tails. The dotted lines are guides to the eyes. The main peaks are deflected quadratically. The deflections are consistent with polarizability of about 5 Å$^3$ per atom. The tails are deflected linearly. The maximum deflections are determined from the location of the baseline intercept and correcting for the natural beam width. The dipole moment of Nb$_{14}$ is about 2 Debye.

FIG. 3(color online). Beam profiles of Nb$_{18}$ at several temperatures. At low temperature, the beam profiles are very asymmetric at $V$=20kV. As the temperature increase, the asymmetry becomes less. At $T$=300K, the deflected beam profile shows only a rigid shift.

FIG. 4(color online). Calculated beam profiles for $TiC_{60}$ at $T=85K$, $v=920m/s$. Lines are calculated by using the CDM by the authors; symbols are calculated by Dugourd *et al*.

FIG. 5(color online). Beam profiles of $Nb_{18}$ at $T=20K$ and $T=313K$ and (a) $V=20kV$, (b)$V=5kV$. $P_{on}$ and $P_{off}$ are the experimental beam profiles with electric field on and off respectively and $P_{on}^{CDM}$ are the profiles with electric field on, calculated by classical dipole model. The calculation uses $\mu=1.3$ Debye, which is found from the extension of the tail (see Fig 2b). $P_{on}^{TC}$ are the profiles with electric field on, calculated by assuming two components, one is polarizable and the other is a dipole moment fixed to the cluster. From the calculation of $P_{on}^{TC}$ of $Nb_{18}$ we find 50% is in the polarizable component both at 5kV and at 20kV, demonstrating that that population of the two components is not field dependent.

FIG. 6(color online). Beam profiles of $Nb_{18}$ at $T=300K$ and $V=20kV$. $P_{on}$ and $P_{off}$ are the experimental beam profiles and $P_{on}^{CDM}$ are the profiles calculated by classical dipole model. (a) using He carrier gas ($v=965m/s$); (b) using Ar carrier gas ($v=435m/s$).

FIG. 7(color online). Experimental beam profiles of $Nb_{28}$ at $T=20K$ and $V=20kV$. $P_{off}$ is the profile with no electric field, $P_{on}$ is the profile with electric field on, and $P_{on}^{Laser}$ is the profile with electric field on, after the cluster beam is illuminated with a 500 nm laser light pulse. Note that laser heating extinguishes the dipole.

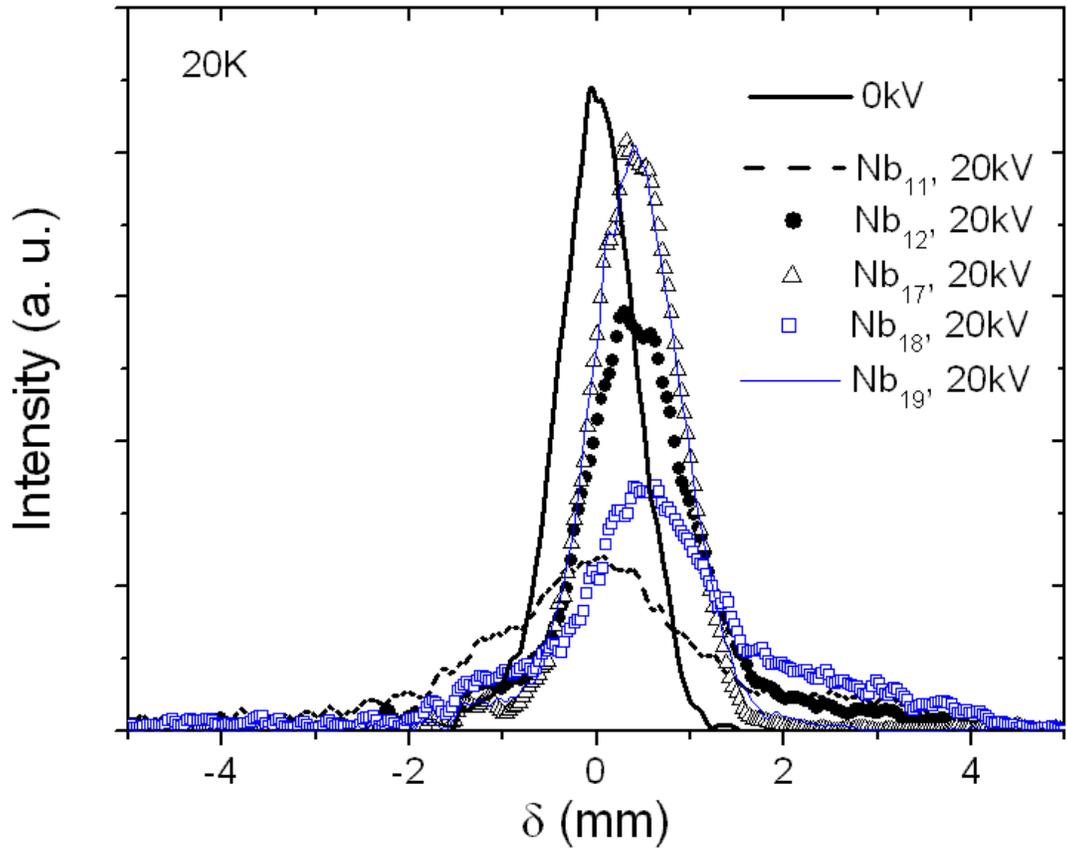

Figure 1

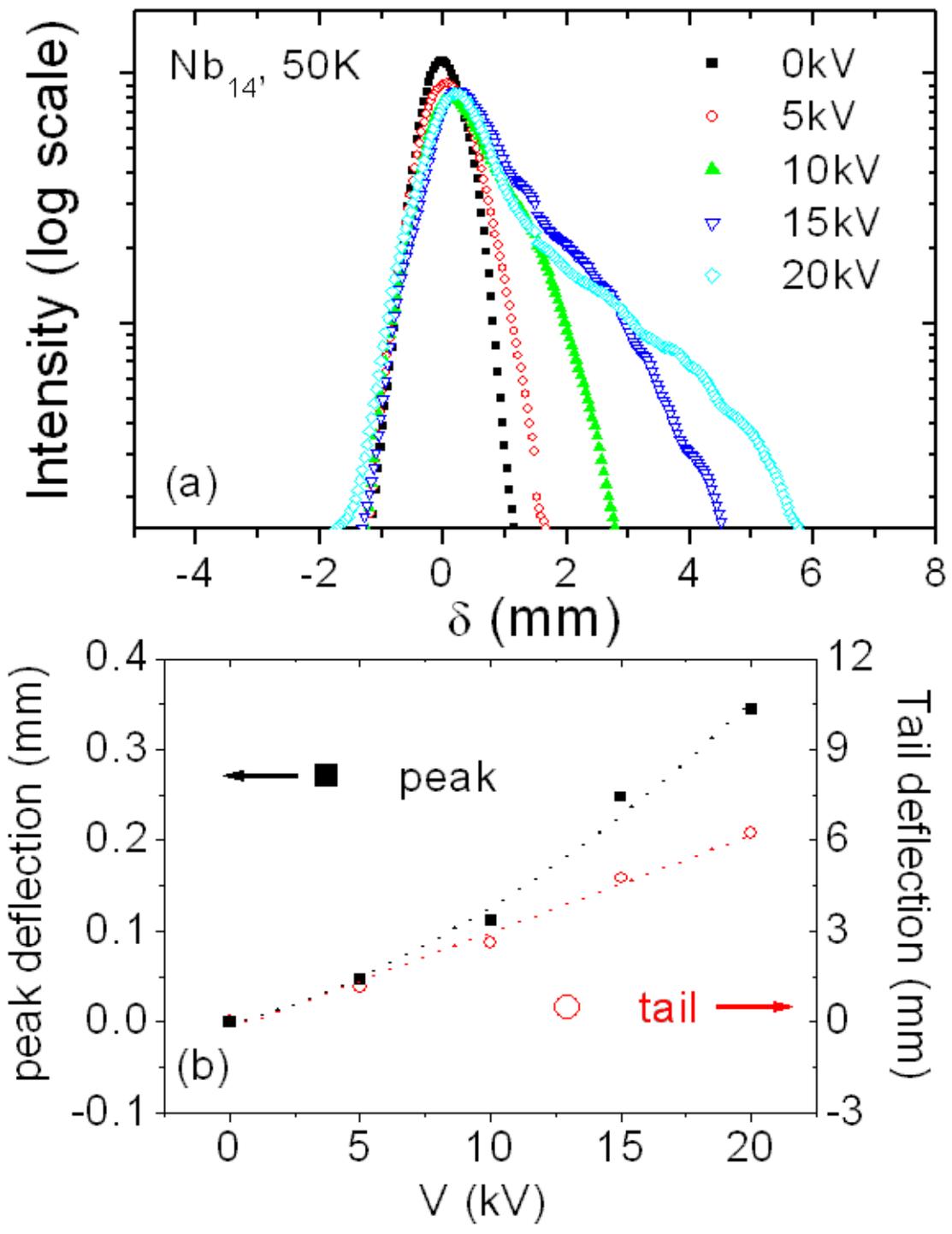

Figure 2

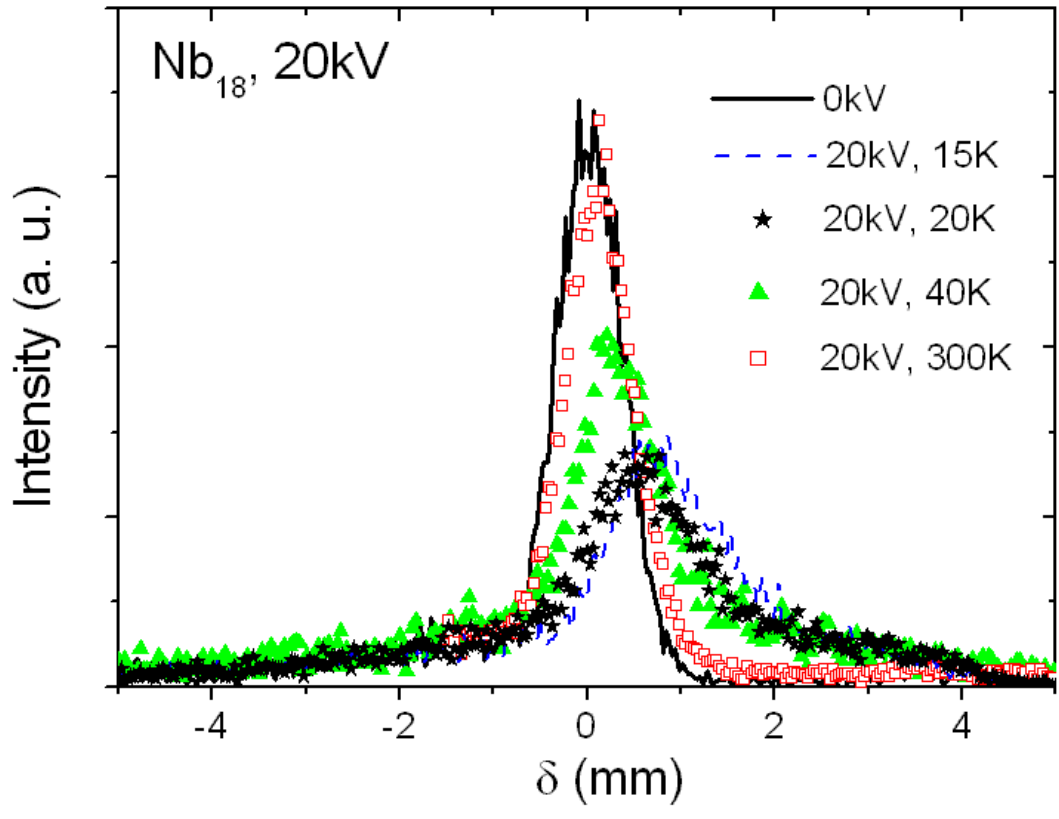

Figure 3

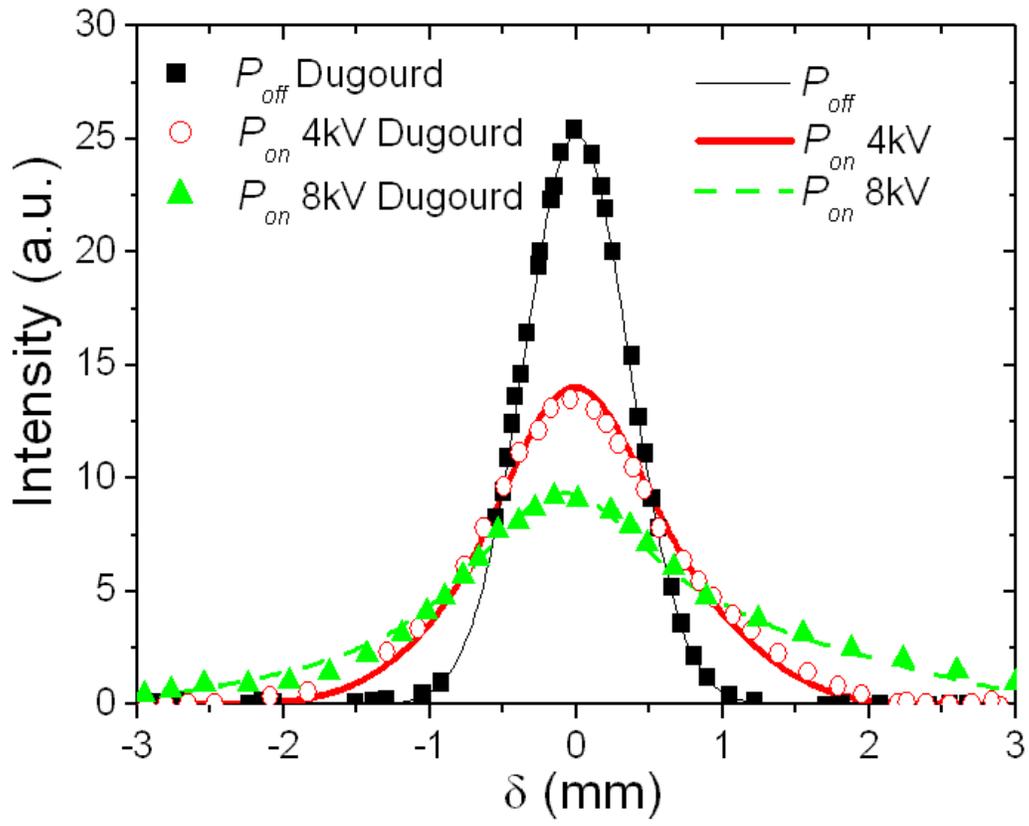

Figure 4

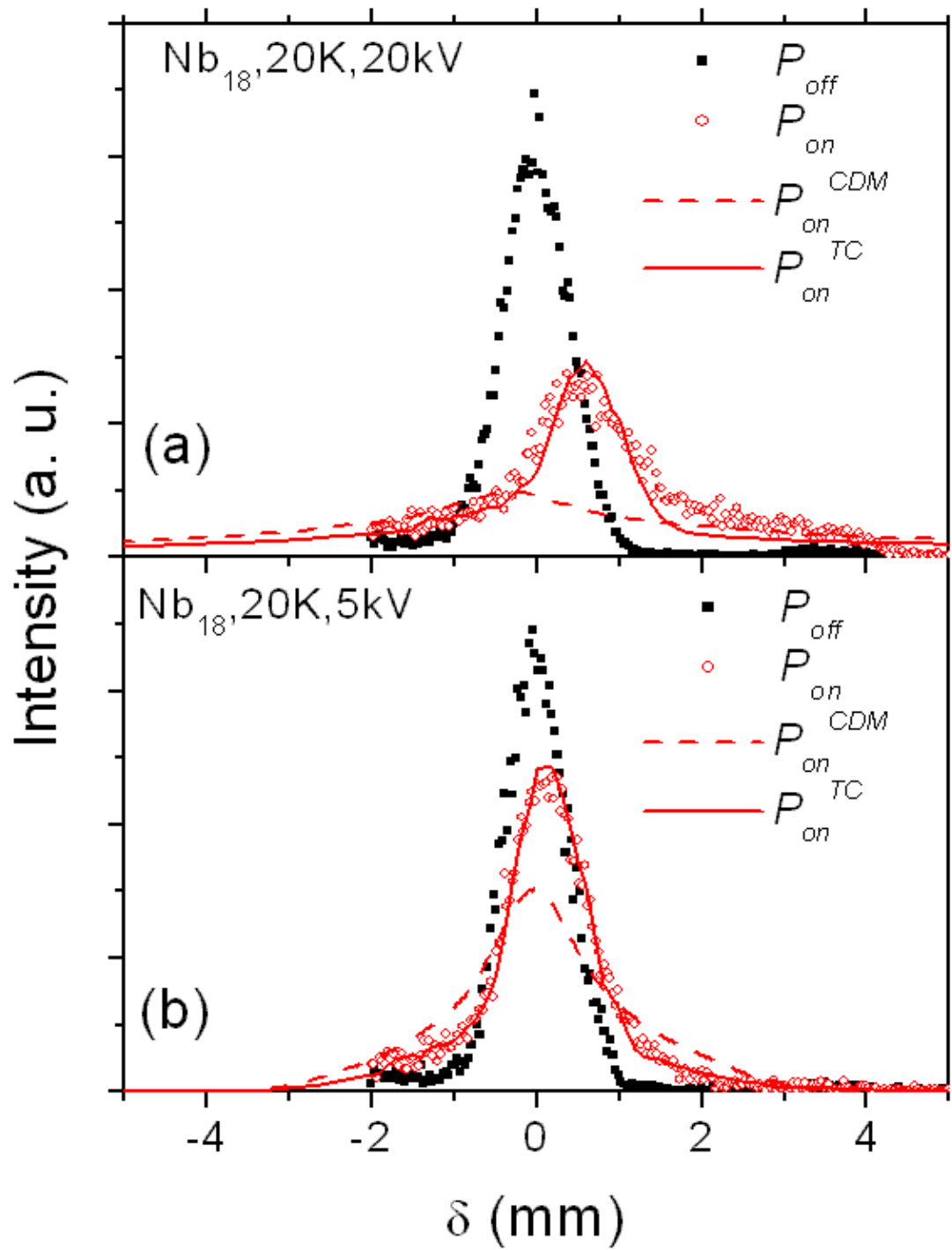

Figure 5

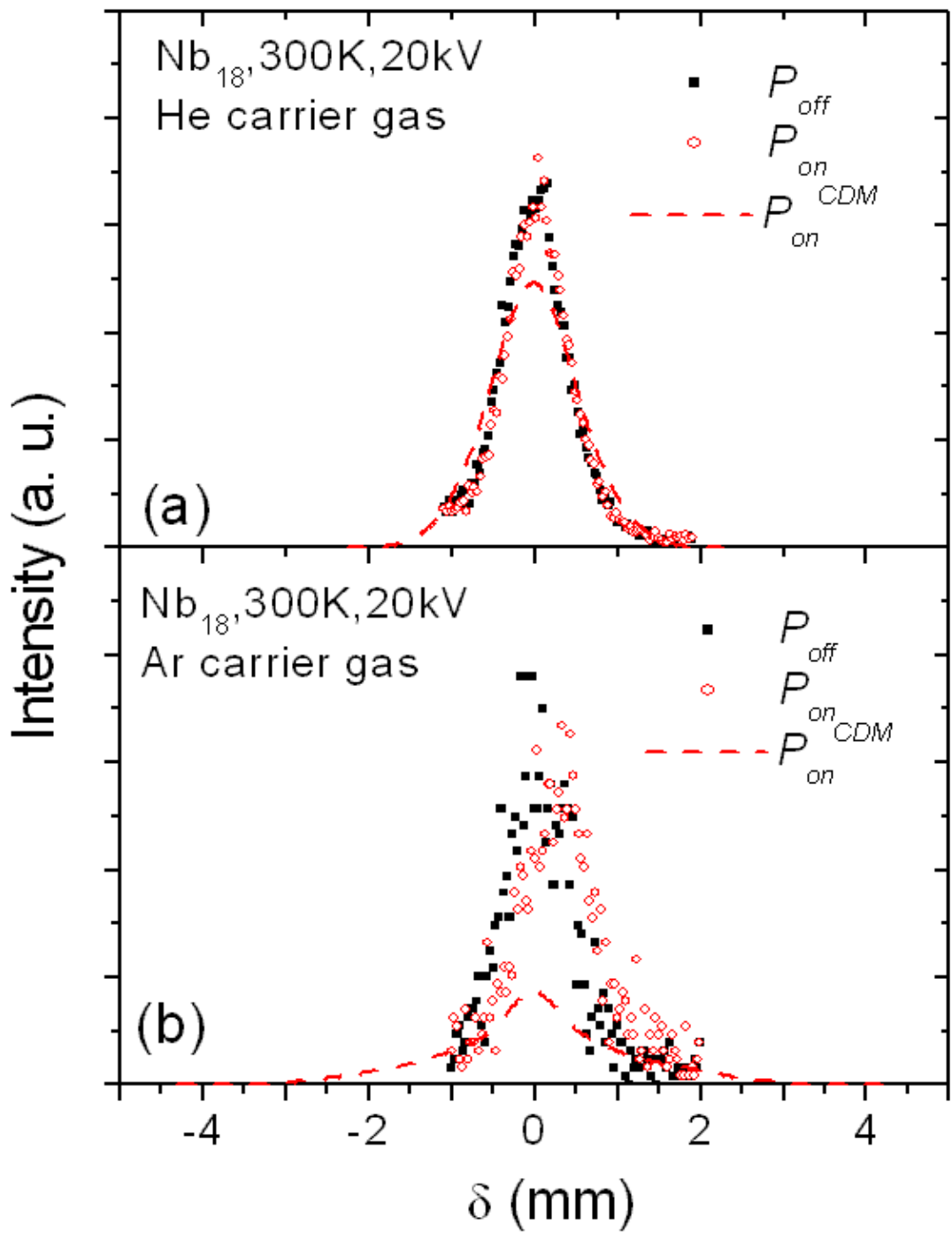

Figure 6

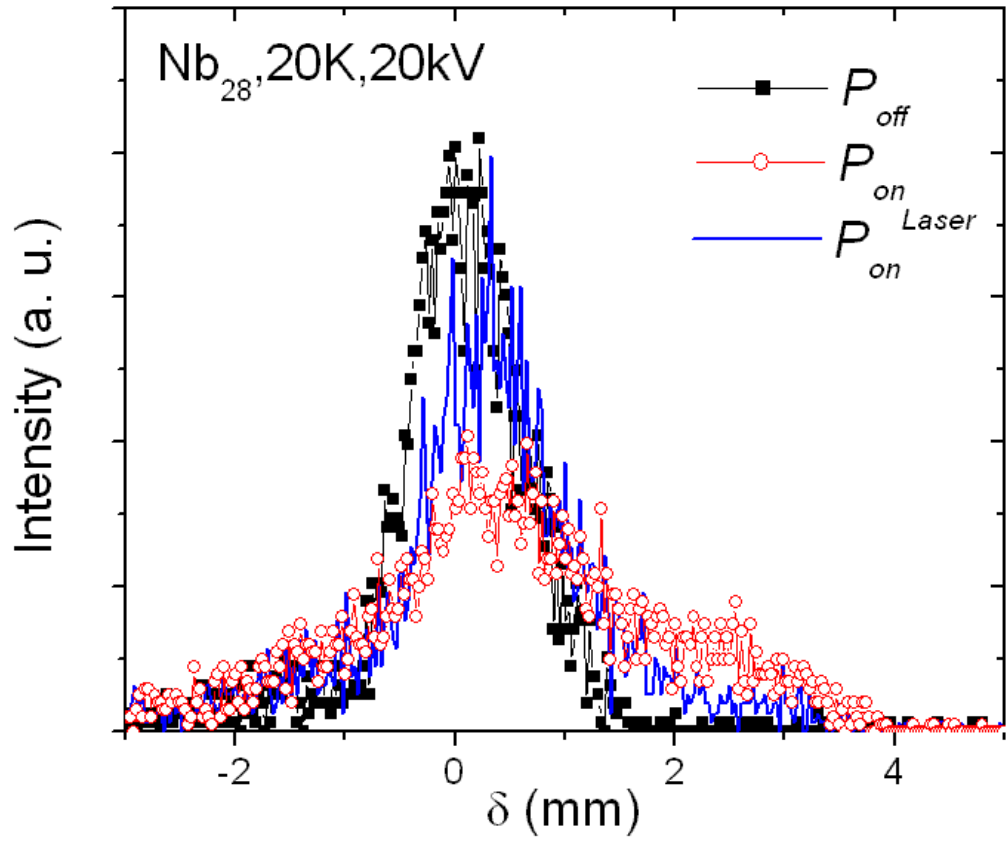

Figure 7